\documentclass[reprint,pre,showpacs,amsmath,amssymb,aps]{revtex4-1}

\usepackage{natbib}
\usepackage{graphicx,color,rotating}
\usepackage[latin1]{inputenc}   
\usepackage{textcomp}
\usepackage{dcolumn}
\usepackage{bm}     
\usepackage[version=3]{mhchem}  

\newcommand{\notop}{{{}_{}}}

\newcommand{\mr}[1]{\ensuremath{\mathrm{#1}}}
\renewcommand{\vec}[1]{\bm{#1}}
\newcommand{\ee}{\mathrm{e}}

\newcommand{\avr}[1]{\big\langle #1 \big\rangle}

\newcommand{\pp}{\partial}       

\newcommand{\nablabf}{\boldsymbol{\nabla}}

\newcommand{\asinh}[1]{ \,\mr{asinh}{\left (#1\right )}}

\newcommand{\vJp}{\vec{J}_+}
\newcommand{\vJm}{\vec{J}_-}
\newcommand{\vJpm}{\vec{J}_{\pm}}

\newcommand{\vJsp}{\vec{J}^{\mathrm{surf}}_{+}}

\newcommand{\Isp}{{I}^{\mathrm{surf}}_
+}
\newcommand{\vJsum}{\vec{J}_{\mathrm{sum}}}
\newcommand{\vJbsum}{\vec{J}^{\mathrm{bulk}}_{\mathrm{sum}}}
\newcommand{\Jbsum}{{J}^{\mathrm{bulk}}_{\mathrm{sum}}}
\newcommand{\vJssum}{\vec{J}^{\mathrm{surf}}_{\mathrm{sum}}}
\newcommand{\vJdif}{\vec{J}_{\mathrm{dif}}}
\newcommand{\vJbdif}{\vec{J}^{\mathrm{bulk}}_{\mathrm{dif}}}
\newcommand{\Jbdif}{{J}^{\mathrm{bulk}}_{\mathrm{dif}}}
\newcommand{\vJsdif}{\vec{J}^{\mathrm{surf}}_{\mathrm{dif}}}

\newcommand{\Isurf}{I_{\mathrm{surf}}}

\newcommand{\vex}{\vec{e}_x}

\newcommand{\ver}{\vec{e}_r}
\newcommand{\vu}{\vec{u}}

\newcommand{\nnn}{\vec{n}}

\newcommand{\zerovec}{\boldsymbol{0}}

\newcommand{\cpm}{c^{{}}_\pm}
\newcommand{\cv}{\bar{c}}  

\newcommand{\rhos}{\rho^{{}}_{\mathrm{s}}}

\newcommand{\cw}{c_{\mathrm{w}}}

\newcommand{\Dpm}{D^\notop_{\pm}}

\newcommand{\kB}{{k^\notop_\mathrm{B}}}

\newcommand{\Penpm}{Pe^0_{\pm}}

\newcommand{\delD}{\delta_{\mathrm{D}}}

\newcommand{\epsw}{\epsilon_{\mathrm{w}}}

\newcommand{\lamD}{\lambda^{{}}_\mathrm{D}}
\newcommand{\blamD}{\bar{\lambda}^{{}}_\mathrm{D}}
\newcommand{\blamG}{\bar{\lambda}^{{}}_\mathrm{G}}
\newcommand{\lamDsqr}{\lambda^{{2}}_\mathrm{D}}
\newcommand{\blamDsqr}{\bar{\lambda}^{{2}}_\mathrm{D}}

\newcommand{\rhoe}{\rho^\notop_\mathrm{el}}

\newcommand{\VT}{V_{\mathrm{T}}}
\newcommand{\ueo}{u_{\mathrm{eo}}}
\newcommand{\udo}{u_{\mathrm{do}}}
\newcommand{\ueou}{u_{\mathrm{eo}}^{\mathrm{u}}}
\newcommand{\udou}{u_{\mathrm{do}}^{\mathrm{u}}}
\newcommand{\ueoup}{u_{\mathrm{eo}}^{\mathrm{up}}}
\newcommand{\udoup}{u_{\mathrm{do}}^{\mathrm{up}}}
\newcommand{\ueoupbnd}{u_{\mathrm{eo,bnd}}^{\mathrm{up}}}
\newcommand{\udoupbnd}{u_{\mathrm{do,bnd}}^{\mathrm{up}}}

\newcommand{\mupm}{{\mu_{\pm}^{{}}{}}}

\newcommand{\phieq}{\phi^{{}}_\mathrm{eq}}

\newcommand{\phib}{\phi^{{}}_\mathrm{bulk}}
\newcommand{\phiGC}{\phi^{{}}_\mathrm{GC}}

\newcommand{\beq}[1]{\begin{equation} \eqlab{#1}}
\newcommand{\eeq}{\end{equation}}
\newcommand{\bsub}{\begin{subequations}}
\newcommand{\esub}{\end{subequations}}
\def\bal#1\eal{\begin{align}#1\end{align}}
\def\bsubal#1\esubal{\bsub \begin{align}#1\end{align} \esub}
\newcommand{\nn}{\nonumber}
\newcommand{\eqlab}[1]{\label{eq:#1}}
\renewcommand{\eqref}[1]{Eq.~(\ref{eq:#1})}
\newcommand{\eqsref}[2]{Eqs.~(\ref{eq:#1}) and~(\ref{eq:#2})}

\newcommand{\figref}[1]{Fig.~\ref{fig:#1}}
\newcommand{\figsref}[2]{Figs.~\ref{fig:#1} and~\ref{fig:#2}}
\newcommand{\figlab}[1]{\label{fig:#1}}

\newcommand{\secref}[1]{Section~\ref{sec:#1}}

\newcommand{\seclab}[1]{\label{sec:#1}}
\newcommand{\tabref}[1]{Table~\ref{tab:#1}}

\newcommand{\tablab}[1]{\label{tab:#1}}

\begin{document}

\title{Concentration polarization, surface currents, and bulk advection in a microchannel}	 

\author{Christoffer P. Nielsen}
\affiliation{Department of Physics, Technical University of Denmark, DTU Physics Building 309, DK-2800 Kongens Lyngby, Denmark}
\email{chnie@fysik.dtu.dk and bruus@fysik.dtu.dk}
\author{Henrik Bruus}
\affiliation{Department of Physics, Technical University of Denmark, DTU Physics Building 309, DK-2800 Kongens Lyngby, Denmark}


\begin{abstract}
\vspace*{-5mm}
\centerline{\small (Submitted to Phys.\ Rev.\ E, 20 August 2014)}

\vspace*{5mm}
We present a comprehensive analysis of salt transport and overlimiting currents in a microchannel during concentration polarization. We have carried out full numerical simulations of the coupled Poisson--Nernst--Planck--Stokes problem governing the transport and rationalized the behaviour of the system. A remarkable outcome of the investigations is the discovery of strong couplings between bulk advection and the surface current; without a surface current, bulk advection is strongly suppressed. The numerical simulations are supplemented by analytical models valid in the long channel limit as well as in the limit of negligible surface charge. By including the effects of diffusion and advection in the diffuse part of the electric double layers, we extend a recently published analytical model of overlimiting current due to surface conduction.

\end{abstract}


\maketitle

\section{Introduction}
\seclab{Intro}

Concentration polarization at electrodes or electrodialysis membranes has been an active field of study for many decades \cite{Eisenberg1954,Porter1972,Nikonenko2010}. In particular, the nature and origin of the so-called overlimiting current, exceeding the diffusion-limited current, has attracted attention.
A number of different mechanisms have been suggested as explanation for this overlimiting current, most of which are probably important for some system configuration or another. The suggested mechanisms include bulk conduction through the extended space-charge region \cite{Rubinstein1979,Manzanares1993}, current induced membrane discharge \cite{Andersen2012}, water-splitting effects \cite{Kharkats1979,Nielsen2014}, electroosmotic instability \cite{Rubinstein2002,Zaltzman2007}, and most recently, electro-hydrodynamic chaos \cite{Druzgalski2013,Davidson2014}.

In recent years, concentration polarization in the context of microsystems has gathered increasing interest \cite{Kim2007,Kim2009,Mani2011,Mani2009,Zangle2009}. This interest has been spurred both by the implications for battery \cite{Linden2002} and fuel cell technology \cite{Tanner1997,Virkar2000,ElMekawy2013} and by the potential applications in water desalination \cite{Kim2010a} and solute preconcentration \cite{Wang2005,Kwak2011,Ko2012}. In microsystems, surface effects are comparatively important, and for this reason their behavior is fundamentally different from bulk systems. For instance, an entirely new mode of overlimiting current enabled by surface conduction, has been predicted by Dydek \emph{et al.}\ \cite{Dydek2011,Dydek2013}, for which the current exceeding the diffusion-limited current runs through the depletion region inside the diffuse double layers screening the surface charges. This gives rise to an overlimiting current depending linearly on the surface charge, the surface-to-bulk ratio, and the applied potential. In addition to carrying a current, the moving ions in the diffuse double layers exert a force on the liquid medium, and thereby they create an electro-diffusio-osmotic flow in the channel. This fluid flow does in turn affect the transport of ions, and the resulting Poisson--Nernst--Planck--Stokes problem has strong nonlinear couplings between diffusion, electromigration, electrostatics, and advection.
While different aspects of the problem can be, and has been, treated analytically \cite{Yaroshchuk2011,Yaroshchuk2011a,Rubinstein2013}, the fully coupled system is in general too complex to allow for a simple analytical description.

In this paper we carry out full numerical simulations of the coupled Poisson--Nernst--Planck--Stokes problem, and in this way we are able to give a comprehensive description of the transport properties and the role of electro-diffusio-osmosis in microchannels during concentration polarization. To supplement the full numerical model, and to allow for fast computation of large systems, we also derive and solve an accurate boundary layer model. We rationalize the results in terms of three key quantities: the Debye length $\blamD$ normalized by the channel radius, the surface charge $\rhos$ averaged over the channel cross-section, and the channel aspect ratio $\alpha$. In the limit of low aspect ratio we derive and verify a simple analytical expression for the current-voltage characteristic, which includes electromigration, diffusion and advection in the diffuse double layers. The overlimiting conductance found in this model is approximately three times larger than the conductance found in Ref.~\cite{Dydek2011}, where diffusion and advection in the diffuse double layers is neglected. In the limit of negligible surface charge the numerical results agree with our previous analytical model \cite{Nielsen2014} for the overlimiting current due to an extended space-charge region.

It has been shown in several papers that reactions between hydronium ions and surface groups can play an important role for the surface charge density and for the transport in microsystems \cite{Behrens2001,Janssen2008,Andersen2011,Jensen2011}. This is especially true in systems exhibiting concentration polarization, as strong pH gradients often occur in such systems.
However, in this work we limit ourselves to the case of constant surface charge density, and defer the treatment of surface charge dynamics to future work.

\section{The model system}
\seclab{model}

Our model system consists of a straight cylindrical microchannel of radius $R$ and length $L$ filled with an aqueous salt solution, which for simplicity is assumed binary and symmetric with valences $Z$ and concentration fields $c_+$ and $c_-$. A reservoir having salt concentration $c_0$ is attached to one end of the channel and a cation-selective membrane to the other end. On the other side of the cation-selective membrane is another reservoir, but due to its relatively simple properties, this part of the system needs not be explicitly modeled and is only represented by an appropriate membrane boundary condition. The channel walls have a uniform surface charge density $\sigma$, which is screened by the salt ions in the liquid over the characteristic length $\lamD$. In \figref{Cylinder_sketch} a sketch of the system is shown. The diffuse double layer adjoining the wall is shown as a shaded (blue) area, and the arrows indicate a velocity field deriving from electro-diffusio-osmosis with back-pressure. We assume cylindrical symmetry and we can therefore reduce the full three-dimensional (3D) problem to a two-dimensional (2D) problem.

\begin{figure}[!t]
    \includegraphics[width=\columnwidth]{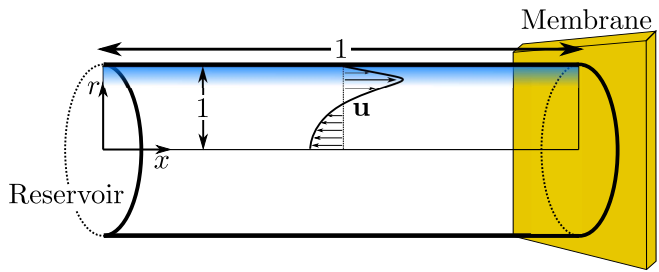}
    \caption{\figlab{Cylinder_sketch} (Color online) A sketch of the axisymmetric 2D system studied in this work. A microchannel of normalized length and radius unity connects a reservoir to the left to a cation-selective membrane to the right. To the right of the membrane is another reservoir, but this part of the system is only modeled through boundary conditions. The diffuse double layer adjoining the wall is shown as a shaded (blue) area and the arrows indicate a velocity field deriving from electro-diffusio-osmosis with back-pressure. }
\end{figure}

\section{Governing equations}
\seclab{Gov_eq}

\subsection{Nondimensionalization}

In this work we use nondimensional variables, which are listed in \tabref{Normalization} together with their normalizations. We further introduce the channel aspect ratio $\alpha$ and the nondimensional gradient operator $\nablabf$,
\bsubal
\alpha &= \frac{R}{L}, \\
\nablabf &= \alpha \vex \pp_x + \ver \pp_r.
\esubal

\begin{table}[!t]
\caption{\tablab{Normalization} Normalizations used in this work. $c_0$ is the reservoir concentration, $Z$ is the valence of the ions, $\VT$ is the thermal voltage, $\kB$ is the Boltzmann constant, $U_0$ is a characteristic electroosmotic velocity, $\epsw$ is the permittivity of water, $\eta$ is the viscosity of water, and $D_+$ and $D_-$ are the diffusivities of the negative and positive ions, respectively.}
\begin{ruledtabular}
\begin{tabular}{lccc}
Variable & Symbol & Normalization  \\ \hline
Ion concentration & $\cpm$ & $c_0$ \\
Electric potential & $\phi$ &  $\VT = \kB T/(Ze)$ \\
Electrochemical potential & $\mupm$ &$\kB T$ \\
Current density & $\vJpm$ & $2D_+c_0/L$ \\
Velocity & $\vu$ & $U_0=\epsw \VT^2/(\eta L)$ \\
Pressure & $p$ & $\eta U_0/R$ \\
Body force density & $\vec{f}$ &  $c_0 \kB T/R$ \\
Radial coordinate & $r$ & $R$ \\
Axial coordinate & $x$ & $L$ \\
Time & $t$ & $R^2/(2D_+)$ \\
\end{tabular}
\end{ruledtabular}
\end{table}

\subsection{Bulk equations}

The nondimensional current density $\vJpm$ of each ionic species of concentration $\cpm$ is given by the electrochemical potentials $\mupm$ and normalization P\' eclet numbers $\Penpm$,
 \bsub
 \bal \eqlab{nondimJ}
 2\alpha \frac{D_+}{\Dpm} \vJpm &= -\cpm\nablabf \mupm+\alpha \Penpm \cpm \vu,\\
 \Penpm &= \frac{L U_0}{\Dpm} = \frac{\epsw \VT^2 }{\eta \Dpm}.
 \eal
For dilute solutions, $\mupm$ can be written as the sum of an ideal gas contribution and the electrostatic potential $\phi$,
 \bal
 \mupm &= \ln(\cpm) \pm \phi. \eqlab{nondimmupm}
 \eal
 \esub
In the absence of reactions, the ions are conserved, and the nondimensional Nernst--Planck equations read
 \begin{align}  \eqlab{nondim_NP}
 \pp_t \cpm = - \alpha \nablabf \cdot  \vJpm.
 \end{align}
The Poisson equation governs $\phi$,
 \begin{align} \eqlab{nondim_Poisson}
 \nabla^2 \phi&= -\frac{1}{2}\frac{R^2}{\lamDsqr} (c_+-c_-) = -\frac{1}{2\blamDsqr} (c_+-c_-),
 \end{align}
where $\blamD = \lamD/R$ is the normalized Debye length, for which $\lamD = \sqrt{\kB T \epsw/(2Z^2e^2c_0) }$ is evaluated for the reservoir concentration $c_0$.
Finally, we have the Stokes and continuity equations governing the velocity field $\vu$, with $x$ and $r$ components $u$ and $v$, and the pressure $p$,
 \bsubal \eqlab{nondim_Stokes}
 \frac{1}{Sc}  \pp_t \vu &= -\nablabf p +\nabla^2 \vu
 + \frac{1}{2\alpha \blamDsqr} \vec{f},\\
 0&= \nablabf \cdot \vu. \eqlab{nondim_Cont}
 \esubal
Here, $Sc = \eta/(\rho D_+)$ is the Schmidt number, and $\vec{f}$ is the body force density acting on the fluid.

\subsection{Thermodynamic forces}
\seclab{Thermodynamic forces}

In an electrokinetic problem, there are essentially two ways of treating the thermodynamic forces driving the ion transport: either the transport is viewed as a result of diffusive and electric forces, or it is viewed as a result of gradients in the electrochemical potential. While the outcome of both approaches is the same, there are some advantages in choosing a certain viewpoint for a specific problem. As the form of \eqref{nondimJ} suggests, we favor the electrochemical viewpoint in many parts of this paper.

\begin{figure}[!t]
    \includegraphics[]{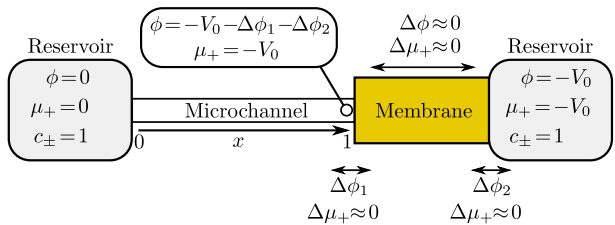}
    \caption{\figlab{Extended_system} (Color online) Sketch of the full physical system including both reservoirs of equal salt concentration. An electric potential difference $V_0$ is applied between the reservoirs, and the changes in electrochemical and electrical potential across the membrane and adjoining Donnan layers are indicated.}
\end{figure}

In \figref{Extended_system}, a sketch of the model system is shown. The system consists of a reservoir on the left, which is connected to another reservoir to the right through a microchannel and an ion-selective membrane. An electric potential difference $V_0$ is applied between the two reservoirs. Typically, the electrical potential drop in the membrane interior is negligible due to the large number of charge carriers in this region, while it varies significantly across the quasi-equilibrium double layers at the membrane interfaces, an effect known as Donnan potential drops \cite{Donnan1995}. In contrast, the cation electrochemical potential is nearly constant across the quasi-equilibrium double layers and thus also across the entire membrane. Unless we want to explicitly model the membrane and the adjoining double layers, it is therefore much more convenient to use the electrochemical potential as control parameter than the electric potential.

Inside the microchannel there are also some advantages of emphasizing the electrochemical potentials. The diffuse double layers screening the surface charges are very close to local equilibrium, meaning that the electrochemical potentials are nearly constant across them. The gradients $\nablabf \mupm$ in electrochemical potentials therefore only have components tangential to the wall, and these components do not vary significantly with the distance from the wall. In contrast, diffusion and electromigration have components in both directions which vary greatly in magnitude through the diffuse double layers.

The electrochemical potentials also offer a convenient way of expressing the body force density $\vec{f}$ from \eqref{nondim_Stokes}. Conventionally, the body force density is set to be the electrostatic force density $-\rhoe \nablabf  \phi = -(c_+-c_-)\nablabf  \phi$. By considering the forces on each constituent we can, however, formulate the problem in a way that is more convenient and better reveals the physics of the problem. The force acting on each particle is minus the gradient of its electrochemical potential. The force density can therefore be written as
 \begin{align}
 \vec{f} &= - c_+ \nablabf \mu_+ - c_- \nablabf \mu_- - \cw \nablabf \mu_{\mr{w}},
 \end{align}
where $\cw \gg \cpm$ and $\mu_{\mr{w}}$ is the concentration and chemical potential of water, respectively. As opposed to $\mupm$ given by the ideal gas \eqref{nondimmupm}, $\mu_{\mr{w}}$ depends linearly on $\cpm$ \cite{LandauStatPhys},
\bsubal
\mu_{\mr{w}} &= -\frac{c_++c_-}{\cw},\\
\vec{f} &= - c_+ \nablabf \mu_+ - c_- \nablabf \mu_- + \nablabf (c_++c_-).  \eqlab{Stokes_force_dens}
\esubal
If we insert the expressions for $\mupm$, the force density  reduces, as it should, to the usual electrostatic force density. It is, however, advantageous to keep the force density on this form, because it reveals the origin of each part of the force. For instance, if we insert a membrane which is impenetrable to ions, only the last term $\nablabf (c_++c_-)$ in the force, can drive a flow across the membrane, because the other forces are transmitted to the liquid via the motion of the ions. It is thus easy to identify $-(c_++c_-)$ as the osmotic pressure in the solution. Inserting \eqref{Stokes_force_dens} for the force $\vec{f}$ in \eqref{nondim_Stokes} and absorbing the osmotic pressure into the new pressure $p' = p - (c_++c_-)$, we obtain
 \begin{align} \eqlab{nondim2_Stokes}
 \frac{1}{Sc}  \pp_t \vu &= -\nablabf p' +\nabla^2 \vu
 - \frac{1}{2\alpha \blamDsqr} \left [c_+ \nablabf \mu_+ + c_-\nablabf \mu_-  \right ].
 \end{align}
We could of course have absorbed any number of gradient terms into the pressure, but we have chosen this particular form of the Stokes equation, due to its convenience when studying electrokinetics. In electrokinetics, electric double layers are ubiquitous, and since the
electrochemical potentials are constant through the diffuse part of the electric double layers, the driving force in \eqref{nondim2_Stokes} is comparatively simple. Also, in this formulation there is no pressure buildup in the diffuse double layers. Both of these features simplify the numerical and analytical treatment of the problem.

\subsection{Boundary conditions}

To supplement the bulk equations (\ref{eq:nondim_NP}), (\ref{eq:nondim_Poisson}), (\ref{eq:nondim_Cont}), and (\ref{eq:nondim2_Stokes}), we specify boundary conditions on the channel walls, at the reservoir, and at the membrane.

At the reservoir $x=0$ we require that the flow $\vu$ is unidirectional along the $x$-axis, and at the channel wall $r=1$ as well as at the membrane surface $x=1$, we impose a no-slip boundary condition
\bsub
\begin{alignat}{2}
\vu &= u\: \vex, \quad &&\text{at }x=0, \\
\vu &=\zerovec, \quad &&\text{at } r=1 \text{ or } x=1.
\end{alignat}
\esub

To find the distribution of the potential $\phi$ at the reservoir $x=0$, we use the assumption of transverse equilibrium in the Poisson equation~(\ref{eq:nondim_Poisson}),
\bsub
\begin{align} \eqlab{nondim_Poisson_bnd}
\frac{1}{r} \pp_r \left (r \pp_r \phi \right )&= \frac{1}{\blamDsqr}\sinh \phi, \quad \text{at }x=0.
\end{align}
Here, $\alpha^2 \pp_x^2 \phi$ in $\nabla^2 \phi$ is neglected in comparison with the large curvature $\frac{1}{r} \pp_r \left (r \pp_r \phi \right )$ in the $r$ direction. The boundary conditions for $\phi$ are a symmetry condition on the cylinder axis $r=0$, and a surface charge boundary condition at the wall $r=1$,
\begin{alignat}{2}
\pp_r \phi &=0, \quad &&\text{at }r=0, \\
\ver \cdot \nablabf \phi &=  -\frac{R\sigma}{\VT \epsw} = \frac{\rhos}{4}\frac{1}{\blamDsqr}, \quad &&\text{at } r=1. \eqlab{Elstat_bnd_nondim}
\end{alignat}
The parameter $\rhos$ is defined as
\begin{align}
\rhos =  -\frac{2 \sigma }{ze c_0 R},
\end{align}
\esub
and physically it is the average charge density in a channel cross-section, which is required to compensate the surface charge density. As explained in Ref.~\cite{Dydek2011}, $\rhos$ is closely related to the overlimiting conductance in the limit of negligible advection.

The boundary conditions for the ions are impenetrable channel walls at $r=1$, and the membrane at $x=1$ is impenetrable to anions while it allows cations to pass,
\bsub
\begin{alignat}{2}
\ver \cdot \vJpm &= 0, &&\quad \text{at } r=1, \\
\vex \cdot \vJm &=0, &&\quad \text{at } x=1.
\end{alignat}
Next to the membrane there is a quasi-equilibrium diffuse double layer, in which the cation concentration increases from the channel concentration to the concentration inside the membrane. Right where this double layer begins, there is a minimum in cation concentration, and we chose this as the boundary condition on the cations. i.e.
\begin{align}
\vex \cdot \nablabf c_+ =0, \quad \text{at }x=1.
\end{align}
\esub

The last boundary conditions relate to $\mupm$ and $p'$. At the reservoir $x=0$, we require transverse equilibrium of the ions, which also leads to the pressure being constant,
\bsub
\begin{alignat}{2}
\mupm &=0, \quad &&\text{at }x=0, \\
p' &=0, \quad &&\text{at }x=0.
\end{alignat}
Finally, as discussed in \secref{Thermodynamic forces}, $\mu_+$ at the membrane $x=1$, is set by $V_0$,
\begin{align} \eqlab{Pot_bnd_nondim}
\mu_+ = -V_0, \quad \text{at }x=1.
\end{align}
\esub

The above governing equations and boundary conditions completely specify the problem and enable a numerical solution of the full Poisson--Nernst--Planck--Stokes problem with couplings between advection, electrostatics, and ion transport. In the remainder of the paper we refer to the model specified in this section as the full model (FULL). See \tabref{Models} for a list of all numerical and analytical models employed in this paper. An issue with the FULL model is that for many systems the computational costs of resolving the diffuse double layers and solving the nonlinear system of equations are prohibitively high. We are therefore motivated to investigate simpler ways of modeling the system, and this is the topic of the following section.

\begin{table}[!t]
\caption{\tablab{Models} The models employed in this paper. }
\begin{ruledtabular}
\begin{tabular}{lll}
Abbreviation & Name & Described in  \\ \hline

FULL & Full model (numerical) & \secref{Gov_eq} \\

BNDF & Boundary layer model, full &  \secref{BL_model} \\

 &  (numerical)&  \\

BNDS & Boundary layer model, slip  & \secref{BL_model} \\

 &  (numerical)&  \\

ASCA & Analytical model, &  \secref{sec_ASCA} \\

& surface conduction-advection &  \\

ASC & Analytical model, & \secref{sec_ASC} \\

 &surface conduction& \\

ABLK & Analytical model,  & \secref{sec_ABLK} \\

 &bulk conduction &  \\

\end{tabular}
\end{ruledtabular}
\end{table}

\section{Boundary layer models}
\seclab{BL_model}

To simplify the problem, we divide the system into a locally electroneutral bulk system and a thin region near the walls comprising the charged diffuse part of the double layer. The influence of the double layers on the bulk system is included via a surface current inside the boundary layer and an electro-diffusio-osmotic slip velocity.

To properly divide the variables into surface and bulk variables, we again consider the electrochemical potentials. In the limit of long and narrow channels the electrolyte is in transverse equilibrium, and the electrochemical potentials vary only along the $x$ direction,
 \bsub
 \begin{align}
 \mupm(x) & = \ln \big[\cpm(x,r)\big] \pm \phi(x,r).
 \end{align}
Since the left-hand side is independent of  $r$, it must be possible to pull out the $x$ dependent parts of $\ln[\cpm(x,r)]$ and $\phi(x,r)$. We denote these parts $\cv_{\pm}(x)$ and $\phib(x)$, respectively, and find
\begin{align}
\mupm(x) & = \ln\left [\cv_{\pm}(x)\right ]+\ln\left [\frac{\cpm(x,r)}{\cv_{\pm}(x)}\right ] \pm \phib(x)\pm \phieq(x,r),
\end{align}
where the equilibrium potential $\phieq(x,r)$ is the remainder of the electric potential, $\phieq=\phi-\phib$. The $r$ dependent parts must compensate each other, which implies a Boltzmann distribution of the ions in the $r$ direction
\begin{align}
\cpm(x,r) = \cv_{\pm}(x)\ee^{\mp \phieq(x,r)}.
\end{align}
The remainder of the electrochemical potentials is then
\begin{align}
\mupm(x) & = \ln\left [\cv_{\pm}(x)\right ]\pm \phib(x).
\end{align}
\esub
For further simplification, we assume that electroneutrality is only violated to compensate the surface charges, i.e. $\cv_+=\cv_-=\cv$. As long as surface conduction or electro-diffusio-osmosis causes some overlimiting current this is a quite good assumption, because in that case the bulk system is not driven hard enough to cause any significant deviation from charge neutrality. For thin diffuse double layers, $\cv$ corresponds to the ion concentration at $r=0$. However, if the Debye length is larger than the radius, $\cv$ does not actually correspond to a concentration which can be found anywhere in the cross-section, and for this reason $\cv$ is often called the virtual salt concentration \cite{Yaroshchuk2011}.

To describe the general case, where transverse equilibrium is not satisfied in each cross-section, we must allow the bulk potential $\phib$ to vary in both $x$ and $r$ direction. Then, however, the simple picture outlined above fails partially, and consequently, we make the ansatz
 \bsub
 \begin{align}
 \cpm(x,r) = \cv(x)\ee^{\mp \phieq(x,r)} + c'(x,r),
 \end{align}
where $c'(x,r)$ accounts for the deviations from transverse equilibrium. Close to the walls, i.e. in or near the diffuse double layer, we therefore have $c'(x,r) \approx 0$. Inserting this ansatz in \eqsref{nondimJ}{nondimmupm} the currents become
 \begin{align}
 2\alpha \frac{D_+}{\Dpm} \vJpm &= -\nablabf \cpm \mp \cpm \nablabf \phi +\alpha \Penpm \cpm \vu
 \\
 &=- \nablabf c' - \ee^{\mp\phieq} \nablabf \cv \mp \cv \ee^{\mp\phieq} \nablabf \phib \nn  \\
 &\quad
 \mp c' \nablabf( \phib+\phieq) + (\cv \ee^{\mp\phieq}+c')\alpha \Penpm \vu \nn  \\
 &= -\nablabf (\cv+c')  \mp (\cv+c') \nablabf \phib \nn  \\
 &\quad
 + (\cv+c')\alpha \Penpm \vu \nn  \\
 &\quad - (\ee^{\mp\phieq}-1) \nablabf \cv \mp \cv(\ee^{\mp\phieq}-1) \nablabf \phib
 \nn \\
 \nn
 &\quad
 \mp c' \nablabf \phieq +\cv(\ee^{\mp\phieq}-1)\alpha \Penpm \vu.
 \end{align}
 \esub
From $\vJpm$, we construct two useful linear combinations, $\vJsum$ and $\vJdif$, as follows,
 \bsub
 \bal
 \alpha \vJsum &=\alpha \left (\vJp + \frac{D_+}{D_- }\vJm\right )
 \\
 &=-\nablabf (\cv+c')  + \alpha \frac{Pe^0_++Pe^0_-}{2} (\cv+c') \vu
 \nn \\
 &\quad
 -(\cosh\phieq-1)\nablabf \cv +\cv \sinh\phieq\nablabf  \phib
 \nn \\
 \nn
 &\quad
 + \alpha \left [ \frac{Pe^0_+}{2}(\ee^{-\phieq}-1)+\frac{Pe^0_-}{2}(\ee^{\phieq}-1) \right ]  \cv  \vu ,
 \eal
 \bal
 \alpha \vJdif &=\alpha \left (\vJp - \frac{D_+}{D_- }\vJm\right )
 \\ &=-(\cv+c') \nablabf \phib  + \alpha \frac{Pe^0_+-Pe^0_-}{2}  (\cv+c')\vu
 \nn \\
 &\quad
 +\sinh\phieq \nablabf \cv \!-\!(\cosh\phieq\!-\!1)\:\cv \nablabf  \phib
 - c' \nablabf \phieq
 \nn \\
 \nn
 &\quad
 + \alpha \left [ \frac{Pe^0_+}{2}(\ee^{-\phieq}-1)-\frac{Pe^0_-}{2}(\ee^{\phieq}-1) \right ]  \cv  \vu.
 \eal
 \esub
The gradient of $\phieq$ is only significant in the diffuse double layer, where by construction $c'\approx 0$. We therefore neglect the $- c' \nablabf \phieq$ term in the expression for $ \vJdif$.
 It is seen that for thin diffuse double layers the terms involving exponentials of $\phieq$ are much larger near the wall than in the bulk. For this reason we divide the currents into bulk- and surface currents
 \bsub
\begin{align}
 \vJsum &= \vJbsum + \vJssum, \\
  \vJdif &= \vJbdif + \vJsdif.
\end{align}
\esub
Here the bulk currents are just the electroneutral parts
\bsub
\eqlab{def_vJbsumdif}
\begin{align}
\alpha \vJbsum &=-\nablabf c  + \alpha Pe^0 c \vu \eqlab{def_vJbsum},\\
\alpha \vJbdif &=-c \nablabf \phib  + \alpha \frac{1-\delD}{1+\delD} Pe^0  c \vu ,
\end{align}
\esub
with $Pe^0 = (Pe^0_++Pe^0_-)/2$ and $\delD = D_+/D_-$. In addition we have introduced the bulk salt concentration
\begin{align}
c(x,r) = \cv(x) + c'(x,r),
\end{align}
which reduces to $\cv(x)$ on the channel walls. We identify the term $-\nablabf c$ in \eqref{def_vJbsum} as the bulk diffusion and the term $\alpha Pe^0 c \vu$ as the bulk advection.
The Nernst--Planck equations corresponding to \eqref{def_vJbsumdif} are
 \bsub
 \begin{align}
 (1+\delD)\pp_t c&= \nabla^2 c - \alpha Pe^0 \nablabf \cdot(c \vu), \eqlab{NP_LENa}\\
 (1-\delD)  \pp_t c&= \nablabf \cdot ( c \nablabf \phib) -\alpha \frac{1-\delD}{1+\delD} Pe^0 \nablabf \cdot(c \vu).\eqlab{NP_LENb}
 \end{align}
 \esub
The surface currents are given by the remainder of the terms. Because the current of anions in the diffuse double layer is very much smaller than the current of cations, the two surface currents $\vJssum$ and $\vJsdif$ are practically identical and equal to the cation current
\begin{align}
 2\alpha \vJsp =& - \cv (\ee^{-\phieq}-1)
 \big[\nablabf \ln(\cv) + \nablabf \phib \big] \nn \\
 &  +\cv (\ee^{-\phieq}-1)\:\alpha Pe^0_+\vu.  \eqlab{Surf_current}
\end{align}
In \figref{BL_sketch}, the division of the system into a surface region and a bulk region is illustrated. The sketch also highlights the distinction between bulk and surface advection.
The first term on the right hand side of \eqref{Surf_current} we denote the surface conduction and the second term the surface advection. Since the surface currents are mainly along the wall we can describe them as scalar currents.
 \begin{align}
 2\alpha  \Isp &= 2\alpha \langle \vec{e}_x \cdot \vJsp \rangle
 \nn \\
 &=-\alpha \cv \avr{ \ee^{-\phieq}-1} \big[\pp_x \ln(\cv)+ \pp_x  \phib\big]
 \nn \\
 &\quad + \alpha Pe^0_+ \cv \avr{(\ee^{-\phieq}-1)\:u} ,
 \eqlab{Isp}
 \end{align}
where the cross-sectional average of any function $f(r)$ is given by the integral $\langle f(r) \rangle = \int_0^1 f(r)\: 2 r \ \mr{d}r$.
The first average is simplified by introducing the mean charge density $\rhos=\langle c_+ -c_- \rangle$ in the channel needed to screen the wall charge. We then find
 \bsubal
 \cv\avr{\ee^{-\phieq}-1} &= \rhos + I_1, \\
 I_1 &=  \cv\avr{\ee^{\phieq} -1},  \eqlab{def_I1}
 \esubal
where $I_1$ is introduced for later use.

\begin{figure}[!t]
    \includegraphics[width=\columnwidth]{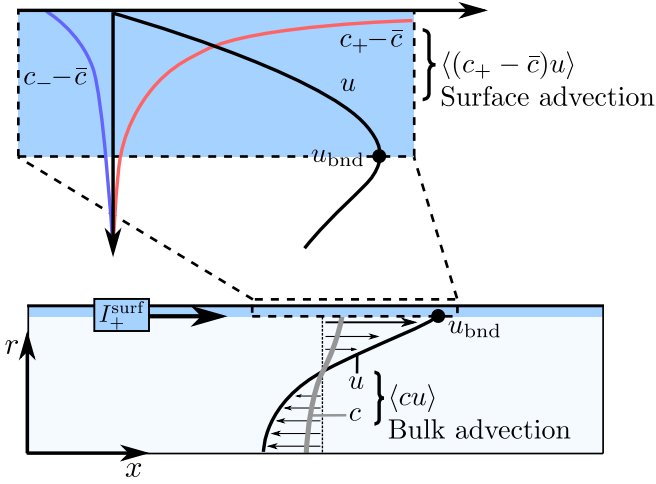}
    \caption{\figlab{BL_sketch} (Color online) Sketch indicating the two regions in the boundary layer model. In the bulk region (lightly shaded) the boundary driven velocity field $u$ (black line), the salt concentration profile $c$ (gray line), and the bulk advection $\langle cu \rangle$ is shown. In the boundary region (shaded and top zoom-in) the excess ion concentrations $\cpm -\cv$, the velocity field $u$, and the surface advection $\langle (\cpm - \cv)u \rangle$ is shown.  }
\end{figure}

Before we proceed with a treatment of the remaining terms in the surface current, there is an issue we need to address: Because of the low concentration in the depletion region, the diffuse double layers are in general not thin in that region. However, the method is saved by the structure of the diffuse double layer in the depletion region. Since the Debye length $\blamD$ is large in the depletion region the negative zeta potential is also large, $-\zeta \gg 1$. The majority of the screening charge is therefore located within the smaller Gouy length $\blamG \ll \blamD$ \cite{Bard2012,Oldham1993}. In \figref{Gouy_plot}, the charge density and the potential are plotted near the channel wall for a system with $\blamD=0.01$ and $\rhos =1$. The charge density is seen to decay on the much smaller length scale $\blamG$ than that of the potential, $\blamD$.
The normalized Gouy-length is given as
 \begin{align}
 \blamG = \frac{\blamD}{\sqrt{c}}\: \asinh{8 \frac{\blamD \sqrt{c}}{\rhos}} \leq 8 \frac{\blamDsqr}{\rhos },
 \end{align}
where the upper limit is a good approximation when $\sqrt{c} \ll \rhos /\blamD$. The boundary layer method is therefore justified provided that
\begin{align}
\blamD \ll 1 \quad \text{or} \quad  8 \frac{\blamDsqr}{\rhos } \ll 1.
\end{align}

\begin{figure}[!t]
    \includegraphics[width=0.9\columnwidth]{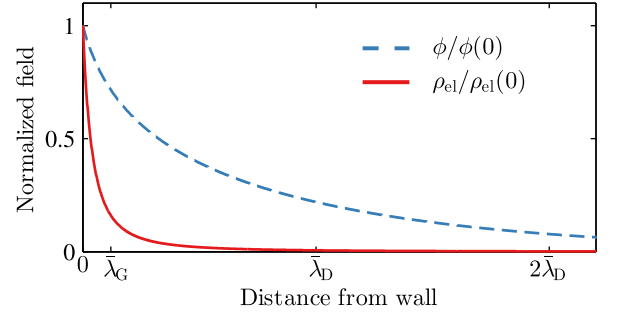}
    \caption{\figlab{Gouy_plot} (Color online) Normalized charge density $\rhoe/\rhoe(0)$ (full) and potential $\phi/\phi(0)$ (dashed) as a function of distance from a charged wall for $\blamD=0.01$ and $\rhos =1$. The Gouy length $\blamG$ and the Debye length $\blamD$ are indicated.}
\end{figure}

To determine the velocity field $u$, we consider the Stokes equation inside the diffuse double layer. In this region the flow is mainly along the wall, and velocity gradients along this direction can be neglected for most cases. The Stokes equation is therefore largely the balance
 \bsub
 \begin{align}
 \frac{1}{Sc}\pp_t u&=
 -\alpha\pp_x p'+\frac{1}{r}\pp_r\big( r \pp_r u \big) \nn \\
 &- \frac{1}{2}\frac{1}{\blamDsqr}\big(c_+\pp_x \mu_++c_-\pp_x \mu_-\big). \eqlab{Stokes_with_time}
 \end{align}
Dimensional analysis shows that the characteristic time scale for the flow inside the diffuse double layer is given by $\blamDsqr /Sc$. For typical systems, where $Sc \gg 1 $ and $\blamDsqr \ll 1$, this time is very much shorter than the bulk diffusion time $\sim 1$, the boundary diffusion time $\sim \blamDsqr$, and the time scale for the bulk flow $\sim 1/Sc$. It is therefore reasonable to neglect the time-derivative term in \eqref{Stokes_with_time}. Assuming Boltzmann distributed ions, $\cpm = \cv \ee^{\mp \phieq}$, and writing out the electrochemical potentials, we obtain
 \bal
 0&= -\alpha\pp_x p'+\frac{1}{r}\pp_r\big( r \pp_r u \big) \nn \\
 &\quad
 + \frac{\cv}{\blamDsqr}\big[ \sinh\phieq \pp_x \phib-\cosh\phieq\pp_x \ln(\cv)\big].
 \eal
Absorbing the bulk diffusive contribution into the new pressure $p''$ we find
 \bal
 0&= -\alpha\pp_x p''+\frac{1}{r}\pp_r\left \{ r \pp_r u \right \} + \frac{\cv}{\blamDsqr}\sinh\phieq \pp_x \phib\nn \\
 &\quad
 - \frac{\cv}{\blamDsqr}\big[\cosh\phieq-\cosh(\phieq(0))\big]\: \pp_x \ln(\cv).
 \eal

This equation is linear in $u$, so we can calculate the electro-osmotic velocity $\ueo$, the diffusio-osmotic velocity $\udo$, and the pressure driven velocity $u_\mr{p}$ individually
 \bal
 u &= \ueo+\udo + u_\mr{p} \\
 &= \ueou \pp_x \phib+\udou \pp_x \ln(\cv) + u_\mr{p},
 \nn \\
 \frac{1}{r}\pp_r\big(r \pp_r \ueou \big)  &=- \frac{\cv}{\blamDsqr}\sinh\phieq ,
 \\
 \frac{1}{r}\pp_r\big( r \pp_r \udou \big)&=  \frac{\cv}{\blamDsqr}\big[\cosh\phieq-\cosh(\phieq(0))\big]\: , \eqlab{udou}
 \\
 \frac{1}{r}\pp_r\big( r \pp_r u_\mr{p} \big)&= \alpha\pp_x p''.
 \eal
 \esub
Here, we also introduced the unit velocity fields $\ueou$ and $\udou$, which both have driving forces of unity. The electroosmotic unit velocity $\ueou$ is found by inserting $\sinh\phieq$ from the Poisson equation and integrating twice
\begin{align}
\ueou = (\zeta-\phieq).
 \end{align}
In the limit $-\zeta \gg 1$, $\cosh\phieq-\cosh(\phieq(0)) \approx -\sinh\phieq$ and the diffusioosmotic unit velocity $\udou$ equals $\ueou$
\begin{align}
\udo = (\zeta-\phieq), \quad \text{for } -\zeta \gg 1.
 \end{align}
 In general, the diffusioosmotic velocity is not as easy to compute, and in practice it is most convenient just to solve \eqref{udou} numerically along with the $\phieq$ problem. The role of the pressure driven velocity fields $u_\mr{p}$ is to ensure incompressibility of the liquid. Rather than dealing with this extra velocity field, we incorporate a pressure driven flow into $\ueou$ and $\udou$ just large enough to ensure no net flux of water through a cross-section
\begin{align}
\ueoup & = \ueou - 2\langle \ueou \rangle (1-r^2), \\
\udoup & = \udou - 2\langle \udou \rangle (1-r^2).
\end{align}
The velocity field can thus be written
 \bal
 u  &= \udoup\: \pp_x \! \ln(\cv )+\ueoup\:\pp_x \phib,
 \eal
with $\langle u \rangle =0$. Using this, we can express the averaged advection term in the surface current \eqref{Isp} as
 \bsubal
 \cv \avr{(\ee^{-\phieq}-1)\:u}
 &=I_2\: \pp_x \phib+I_3\:\pp_x\!\ln(\cv ),\\
 \eqlab{def_I2_and_I3}
 I_2 & =  \cv \avr{(\ee^{-\phieq}-1)\:\ueoup }, \\
 I_3 & = \cv \avr{(\ee^{-\phieq}-1)\:\udoup }.
\esubal
The surface current can then be written as
 \bal
 2\alpha \Isp &= -\alpha (\rhos +I_1)\big[\pp_x \phib+\pp_x \ln(\cv)\big]
 \nn \\
 &\quad
 + \alpha Pe^0_+\big[I_2\pp_x \phib+I_3\pp_x \ln(\cv)\big].
 \eqlab{I_surf}
 \eal
The current into the diffuse double layer from the bulk system is
 \bal \eqlab{BCIsurf}
 \vec{n} \cdot \vJp = \frac{1}{2} \alpha \pp_x \Isp,
 \eal
where the factor of a half comes from the channel cross-section divided by the circumference.
Rather than resolve the diffuse double layers we can therefore include their approximate influence through the boundary condition \eqref{BCIsurf}.

\begin{figure}[!t]
    \includegraphics[width=0.9\columnwidth]{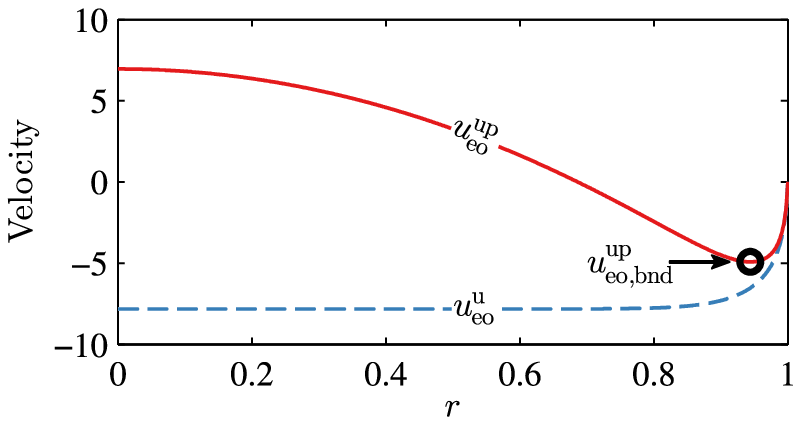}
    \caption{\figlab{ueou_plot} (Color online) The electroosmotic flow $\ueou$ and the electroosmotic flow $\ueoup$ with backpressure for $\rhos =10$ and $\blamD = 0.05$. The effective boundary velocity $\ueoupbnd$ is also indicated. }
\end{figure}

In the locally electroneutral bulk system the Stokes and continuity equations become
\bsub \eqlab{NS_LEN}
\begin{align}
\frac{1}{Sc}  \pp_t \vu &= -\nablabf p' +\nabla^2 \vu ,\\
 0&= \nablabf \cdot \vu.
\end{align}
\esub
The effects of electroosmosis and diffusioosmosis are included via a boundary condition at the walls
 \begin{align}
 \vu &= u_\mr{bnd} \vec{e}_x
 = \left [\ueoupbnd\pp_x \phib+\udoupbnd\:\pp_x \ln(\cv)\right ]\vec{e}_x,
 \nn \\ & \text{at }r=1,   \eqlab{ubnd}
 \end{align}
where $\ueoupbnd$ and $\udoupbnd$ are the minimum values of $\ueoup$ and $\udoup$, i.e. the velocity at the point where the back-pressure driven flow becomes significant. In \figref{ueou_plot} some of the discussed velocity fields are illustrated for $\rhos =10$ and $\blamD = 0.05$. Note that $\pp_x \phib$ and $\pp_x \ln(\cv)$ will most often be negative, so the actual velocities in the channel differ from the plotted velocities with a sign and a numeric factor.

In the remainder of the paper we refer to the model  developed in this section as the full boundary layer (BNDF) model. We also introduce the slip boundary layer (BNDS) model, in which the bulk couples to the boundary layers only through a slip velocity, while the boundary condition~(\ref{eq:BCIsurf}) for the normal current is substituted by  $\nnn\cdot\vJp = 0$. In other words, the BNDS and BNDF models are identical, except the BNDS model does not include the surface current. These models are listed in \tabref{Models} along with the other models of the paper.

\section{Analysis} \seclab{Analysis}

\subsection{ Scaling of bulk advection}
\seclab{Scalingofbulkadv}

To estimate the influence of bulk advection we consider the bulk current for a system in steady state
 \begin{align}
 \alpha \vJbsum &=-\nablabf c  + \alpha Pe^0 c \vu.
 \end{align}
The average of this current in the $x$-direction is
 \begin{align}
 \Jbsum &= \langle \vec{e}_x \cdot \vJbsum \rangle = - \pp_x \langle c\rangle + Pe^0 \langle c u \rangle. \eqlab{Jbsum_sim}
 \end{align}
Since the membrane blocks the flow in one end, the net flow $\langle u \rangle $ in a channel cross-section is zero and thus $\avr{ \cv(x) u} = \cv(x) \avr{u} =0$, which leads to,
 \bal
 \avr{ c u} =  \avr{[\cv(x) +c'(x,r)] u } =  \avr{c'(x,r) u}.
 \eal
Now, the source of the deviation $c'$ between $\cv$ and $c$ is the flow itself. In steady state, the dominant balance in \eqref{NP_LENa} is
 \bsub
 \bal
 \frac{1}{r} \pp_r(r \pp_r c) &\approx \alpha^2 Pe^0 \pp_x (c u),
 \eal
so $c'$ must scale as
 \bal
 c' \sim \alpha^2 Pe^0 \pp_x (\cv u),
 \eal
which upon insertion in \eqref{Jbsum_sim} yields
 \bal
 \Jbsum \sim - \pp_x \avr{c} + (\alpha Pe^0)^2 \avr{ \pp_x (\cv u) u}. \eqlab{Jbsum_sim_scaling}
 \eal
 \esub
This approximative expression reveals an essential aspect of the transport problem: With the chosen normalization neither the velocity, the diffusive current, the electromigration current, nor the surface current depend on the aspect ratio $\alpha$. The only term depending on $\alpha$, is the bulk advection, and we see that for long slender channels ($\alpha \ll 1$) bulk advection vanishes, whereas it can be significant for short broad channels ($\alpha \gg 1$).

\subsection{Local equilibrium models for small $\vec{\alpha}$}

In the limit $\alpha \ll 1$, where bulk advection has a negligible effect, we can derive some simple analytical results. There the bulk concentration $c(x,r)$ equals the virtual concentration $\cv(x)$, and the area-averaged bulk currents are
\bsub
\begin{align}
 \Jbsum &=- \pp_x  \cv(x) ,\\
 \Jbdif &=- \cv(x)  \pp_x  \phib (x).
\end{align}
\esub
In steady-state, these currents are equal and can only change if there is a current into or out of the boundary layer. The conserved current $J_+$ is therefore
\begin{align}
 J_+ &=- \pp_x  \cv(x)+ \Isurf \nn \\
& = -\cv(x)  \pp_x  \phib (x)+ \Isurf.  \eqlab{Conserved_current}
\end{align}
It is readily seen that $\cv=\ee^{\phib}$ is a solution to the equation. To proceed we need expressions for the integrals $I_1$, $I_2$ and $I_3$.

Initially, we neglect advection in the boundary layer as well and this leaves us with the equation
\begin{align}
J_+ = - \ee^{\phib} \pp_x \phib -\frac{1}{2}(\rhos +I_1)2\pp_x \phib.
\end{align}
If the Debye--H\"{u}ckel limit is valid in the diffuse double layer, we can make the approximations
 \bsubal
 \rhos &= c \avr{\ee^{-\phieq}-\ee^{\phieq}} \approx - 2c \avr{\phieq},\\
 I_1 &=c \langle \ee^{\phieq}-1 \rangle \approx c\avr{\phieq}
 \approx -\frac{1}{2}\:\rhos,
 \esubal
in which case $J_+$ reduces to the expression in Ref.~\cite{Dydek2011},
 \bal
 J_+ = -\left (\ee^{\phib}+\frac{\rhos}{2} \right )\pp_x \phib.
 \eal
If, on the other hand, the diffuse double layer is in the strongly nonlinear regime, then the surface charge is compensated almost entirely by cations and to a good approximation
\begin{align}
I_1 \approx 0.
\end{align}
In that limit the current is
\begin{align} \eqlab{OurDydek}
J_+ = -\left (\ee^{\phib}+\rhos \right )\pp_x \phib,
\end{align}
 i.e.\ the overlimiting conductance is twice the conductance found in Ref.~\cite{Dydek2011}. Since the Debye length is large in the depletion region, we have $-\zeta \gg 1$, and the diffuse double layer is in the strongly nonlinear regime. Surface conduction is mainly important in the depletion region, so for most parameter values \eqref{OurDydek} is a fairly accurate expression for the current.

We now make a more general treatment, which is valid when the characteristic dimension of the diffuse double layer is much smaller than the channel curvature. In that limit we can approximate the equilibrium potential with the Gouy--Chapman solution,
 \bsubal
 \phiGC &=
 4\mathrm{atanh}\bigg\{\tanh\bigg[\frac{\zeta}{4}\bigg]\:
 \mathrm{exp}\bigg[-\sqrt{\cv}\:\frac{y}{\blamD}\bigg]\bigg\}, \\
 \zeta &= -2\mathrm{asinh}\bigg[\frac{\rhos}{4d\blamD\sqrt{\cv}}\bigg] \approx -2\ln\bigg[\frac{\rhos}{2d\blamD\sqrt{\cv}}\bigg],
 \esubal
where the last approximation is valid for $-\zeta \gtrsim 2$. In the following we assume that we are in this limit. The parameter $d$ is the ratio of circumference to area of the channel ($d=2$ for a cylindrical channel). Using the Gouy-Chapman solution we find an expression for $I_1$
 \bal
 I_1&=\cv\avr{\ee^{\phieq} -1}
 \approx d\cv \int_0^{\infty} \ee^{\phieq} -1 \ \mr{d}y
 \nn \\
 &= -2 d \blamD \sqrt{\cv}\:\big(1-\ee^{\frac{1}{2}\zeta}\big) \nn \\
 &\approx  4d^2\frac{\blamDsqr}{\rhos}\:\cv -2d\blamD \sqrt{\cv}.
 \eal

In the limit of large potentials, $-\phiGC \gg 1$ we can approximate $\cosh\phiGC \approx -\sinh\phiGC$ and obtain
 \bsubal
 \ueo &= (\zeta-\phiGC)\:\pp_x \phib, \\
 \udo &\approx (\zeta-\phiGC)\:\pp_x \ln(c).
 \esubal
From this we find
 \begin{align}
 &\avr{\cv(\zeta-\phiGC)(\ee^{-\phiGC}-1)} \nn \\
 &\approx d \cv  \int_0^{\infty} (\zeta-\phiGC)(\ee^{-\phiGC}-1) \ \mr{d}y \nn \\
 &= 4 d \blamD \sqrt{\cv} \left (1 -\frac{1}{2}\zeta - \ee^{-\frac{1}{2}\zeta} \right ) \nn \\
 &\approx   4d\blamD \sqrt{\cv} +4d \blamD \sqrt{\cv} \ln\left (\frac{\rhos}{2d\sqrt{\cv}\blamD}\right )-2\rhos. \eqlab{analI2}
 \end{align}
Inserting \eqref{analI2} in \eqsref{I_surf}{Conserved_current} we obtain,
\bsub

 \bal
 J_+ &= - \ee^{\phib} \pp_x \phib \nn \\
 &\quad -\left (\rhos +4 d^2 \frac{\blamDsqr}{\rhos}\ee^{\phib}-2d\blamD \ee^{\frac{1}{2}\phib} \right ) \pp_x \phib\nn  \\
 &\quad - Pe^0_+ \bigg [2\rhos -4d\blamD \ee^{\frac{1}{2}\phib}\nn  \\
 &\quad -4d\blamD \ee^{\frac{1}{2}\phib} \ln\left (\frac{\rhos}{2d} \frac{\ee^{-\frac{1}{2}\phib}}{\blamD}\right ) \bigg ] \pp_x \phib.   \eqlab{full_analytical1}
 \eal
Integration of this expression with respect to $x$ leads to
 \bal
 J_+ x &= \left (1+4d^2\frac{\blamDsqr}{\rhos}\right )
 \big(1- \ee^{\phib}\big)  -\rhos\big(1+2Pe^0_+\big)\phib  \nn \\
 &\quad
 - 4d\blamD\big(1+2Pe^0_+\big)\big( 1-\ee^{\frac{1}{2}\phib} \big)  \nn \\
 &\quad	
 -8d Pe^0_+\blamD  \bigg\{ \left (1+\ln\left [\frac{\rhos}{2d\blamD}\right ] \right )( 1-\ee^{\frac{1}{2}\phib}) \nn \\
 &\qquad \qquad \qquad \quad
 +\frac{1}{2}\phib\:\ee^{\frac{1}{2}\phib} \bigg\}. \eqlab{full_analytical}
\eal
\esub
\subsection{Analytical surface conduction and surface advection (ASCA) model}  \seclab{sec_ASCA}

For $\blamD\ll 1$, the leading order behaviour of \eqsref{full_analytical1}{full_analytical} is
\bsub
\begin{align}
J_+  &= -\ee^{\phib}\pp_x \phib  -\rhos(1+2Pe^0_+)\pp_x\phib \eqlab{sim_surf_cond_adv_phib1} \\
J_+ x &= 1- \ee^{\phib}  -\rhos(1+2Pe^0_+)\phib. \eqlab{sim_surf_cond_adv_phib}
\end{align}
In \eqref{sim_surf_cond_adv_phib1} it is seen that the bulk conductivity $\ee^{\phib}$ varies with the electric potential, whereas the surface conductivity $\rhos(1+2Pe^0_+)$ is constant. At $x=1$, the boundary condition for the potential is $\mu_+ = \ln(\cv)+\phib = 2 \phib = -V_0$, and from \eqref{sim_surf_cond_adv_phib} we obtain the current-voltage relation
\begin{align}
J_+ &= 1- \ee^{-\frac{1}{2}V_0}  +\rhos\left (\frac{1}{2}+Pe^0_+\right )V_0.  \eqlab{sim_surf_cond_adv}
\end{align}
\esub
While this expression was derived with a cylindrical geometry in mind, it applies to most channel geometries.
The only requirement is that the local radius of curvature of the channel wall is much larger than the Gouy length $\blamG$, so that the potential is well approximated by the Gouy-Chapman solution.

This analytical model is called the surface conduction-advection (ASCA) model. As shown in \secref{NumRes}, it is very accurate in the limit of long slender channels, $\alpha \ll 1$.

\subsection{Analytical surface conduction (ASC) model}
 \seclab{sec_ASC}

For a system with a Gouy length on the order of unity, the screening charges are distributed across the channel in the depletion region. Advection therefore transports approximately as many cations towards the membrane as away from the membrane, and there is no net effect of surface advection. In this limit,  \eqref{sim_surf_cond_adv} reduces to the pure surface conduction expression
\begin{align}
J_+ &= 1- \ee^{-\frac{1}{2}V_0}  +\frac{\rhos}{2}V_0, \eqlab{sim_surf_cond}
\end{align}
which we refer to as the analytical surface conduction (ASC) model.

\subsection{Analytical bulk conduction (ABLK) model}
 \seclab{sec_ABLK}

In the limit of low surface charge and high $\blamD$, neither surface conduction nor advection matter much. In that limit the dominant mechanism of overlimiting current is bulk conduction through the extended space-charge region (ESC). This effect is not captured by the derived boundary layer model, since it assumes local electroneutrality. The development of an extended space-charge region can, however, be captured in an analytical 1D model, and from Ref.~\cite{Nielsen2014} we have the limiting expression
 \begin{align}
 -V_0 = \mu_+(1) \approx -\frac{2\sqrt{2}}{3}
 \frac{ (J_+-1)^{3/2}}{\alpha\blamD J_+} + 2\ln(\alpha \blamD), \eqlab{ESC_cond_asympt}
 \end{align}
giving the overlimiting current-voltage characteristic due to conduction through the extended space-charge region. Expressions, which are uniformly valid both at under- and overlimiting current, are also derived in our previous work Ref.~\cite{Nielsen2014}, but since these are rather lengthy we will not show them here. We refer to the full model from Ref.~\cite{Nielsen2014} as the analytical bulk conduction (ABLK) model, see \tabref{Models}.

\section{Numerical analysis}
\seclab{numerics}

\subsection{Numerical implementation}
The numerical simulations are carried out in the commercially available finite element software \textsc{COMSOL Multiphysics} ver.~4.3a. Following Gregersen \emph{et al.}~\cite{Gregersen2009}, the governing equations of the FULL, BNDF, and BNDS models are rewritten in weak form and implemented in the mathematics module of \textsc{COMSOL}. To improve the numerical stability of the problem we have made a change of variable, so that the logarithm of the concentration fields have been used as dependent variables instead of the concentration fields themselves. The cross-sectional averages $I_1$, $I_2$, and $I_3$ (\eqsref{def_I1}{def_I2_and_I3}) as well as the slip velocity (\eqref{ubnd}) are calculated and tabulated in a separate model.

\begin{table}[!t]
\caption{\tablab{Parameters} Parameters and their values or range of values. The Schmidt number is irrelevant since we are considering steady-state problems. To simplify the analysis, $Pe^0$ and $\delD$ are fixed.}
\begin{ruledtabular}
\begin{tabular}{lcl}
Parameter & Symbol & Value/Range  \\ \hline

Schmidt number & $Sc$ & N/A \\

Normalization P\' eclet number & $Pe^0$ &  $0.235$ \\

Diffusivity ratio & $\delD$ &$1$ \\

Aspect ratio & $\alpha$ & 0.01--0.2 \\

Normalized Debye length & $\blamD$ & 0.0001--0.1 \\

Average surface charge density & $\rhos$ & 0.001--1 \\

Bias voltage & $V_0$ &  0--100 \\

\end{tabular}
\end{ruledtabular}
\end{table}

In the theoretical treatment we found seven dimensionless numbers, which govern the behaviour of the system. These are the Schmidt number $Sc$, the normalization P\' eclet number $Pe^0$, the diffusivity ratio $\delD$, the aspect ratio $\alpha$, the normalized Debye length $\blamD$, the cross-sectionally averaged charge density $\rhos$, and the applied bias voltage $V_0$. In the numerical simulations, we only consider steady state problems, so $Sc$ does not matter for the results. To further limit the parameter space, we have chosen fixed and physically reasonable values for a few of the parameters.
The ionic diffusivities are assumed to be equal, i.e. $\delD=1$. For a solution of potassium chloride with $D_{\mr{K}^+} = 1.96\ \mr{m}^2/\mr{s}$ and $D_{\mr{Cl}^-} = 2.03\ \mr{m}^2/\mr{s}$, this is actually nearly the case. The normalization P\' eclet number is set to $Pe^0=0.235$, which is a realistic number for potassium ions in water at room temperature. This leaves us with four parameters, $\alpha$, $\blamD$, $\rhos$ and $V_0$, which govern the system behaviour. We mainly present our results in the form of $I$-$V$ characteristics, i.e. sweeps in $V_0$, since the important features of the transport mechanisms can most often be inferred from these. We vary the other parameters as follows: the aspect ratio $\alpha$ takes on the values $\{0.01,0.05,0.1,0.2\}$, the normalized Debye length $\blamD$ takes the values $\{0.0001,0.001,0.01,0.1\}$, and the averaged charge density $\rhos$ takes the values $\{0.001,0.01,0.1,1\}$. The parameters and their values or range of values are listed in \tabref{Parameters}. The $\blamD =0.0001$ systems are only solved in the BNDF model, since a full numerical solution with resolved diffuse double layers is computationally costly in this limit $\blamD \ll 1$. The boundary layer model is very accurate in the small $\blamD$ limit, so the lack of a full numerical solution for $\blamD=0.0001$ is not a concern.

To verify the numerical scheme we have made comparisons with known analytical results in various limits and carried out careful mesh convergence analyses for selected sets of parameter values.

\subsection{Parameter dependence of $\vec{I-V}$ characteristics} \seclab{NumRes}

The results of the simulations are presented in the following way: For each $\alpha$ value a $(\blamD,\rhos)$ grid is made, and in each grid point is shown the corresponding $I$-$V$ characteristic. The $I$-$V$ characteristics obtained from the simulations are supplemented with relevant analytical results. To aid in the interpretation of the results, \figref{IV_plots_rationalize} shows the trends we expect on the basis of the governing equations and our analysis. Surface conduction and surface advection is expected to increase with $\rhos$, bulk advection is expected to increase with $\rhos$ and $\alpha$ and decrease with $\blamD$. Bulk conduction through the extended space-charge region is expected to increase with $\alpha \blamD$.

\begin{figure}[!t]
    \includegraphics[width=0.8\columnwidth]{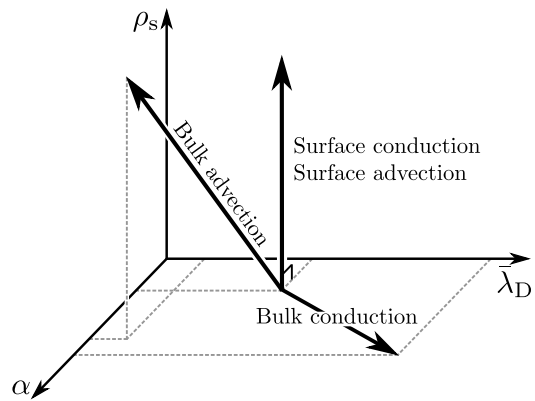}
    \caption{\figlab{IV_plots_rationalize} Directions of increase of the various mechanisms of overlimiting current. Bulk advection increases with $\alpha$ and $\rhos$ and decreases with $\blamD$. Surface conduction and surface advection increases with $\rhos$, and bulk conduction through the ESC increases with $\alpha \blamD$.}
\end{figure}

\begin{figure*}[!t]
    \includegraphics[width=17cm]{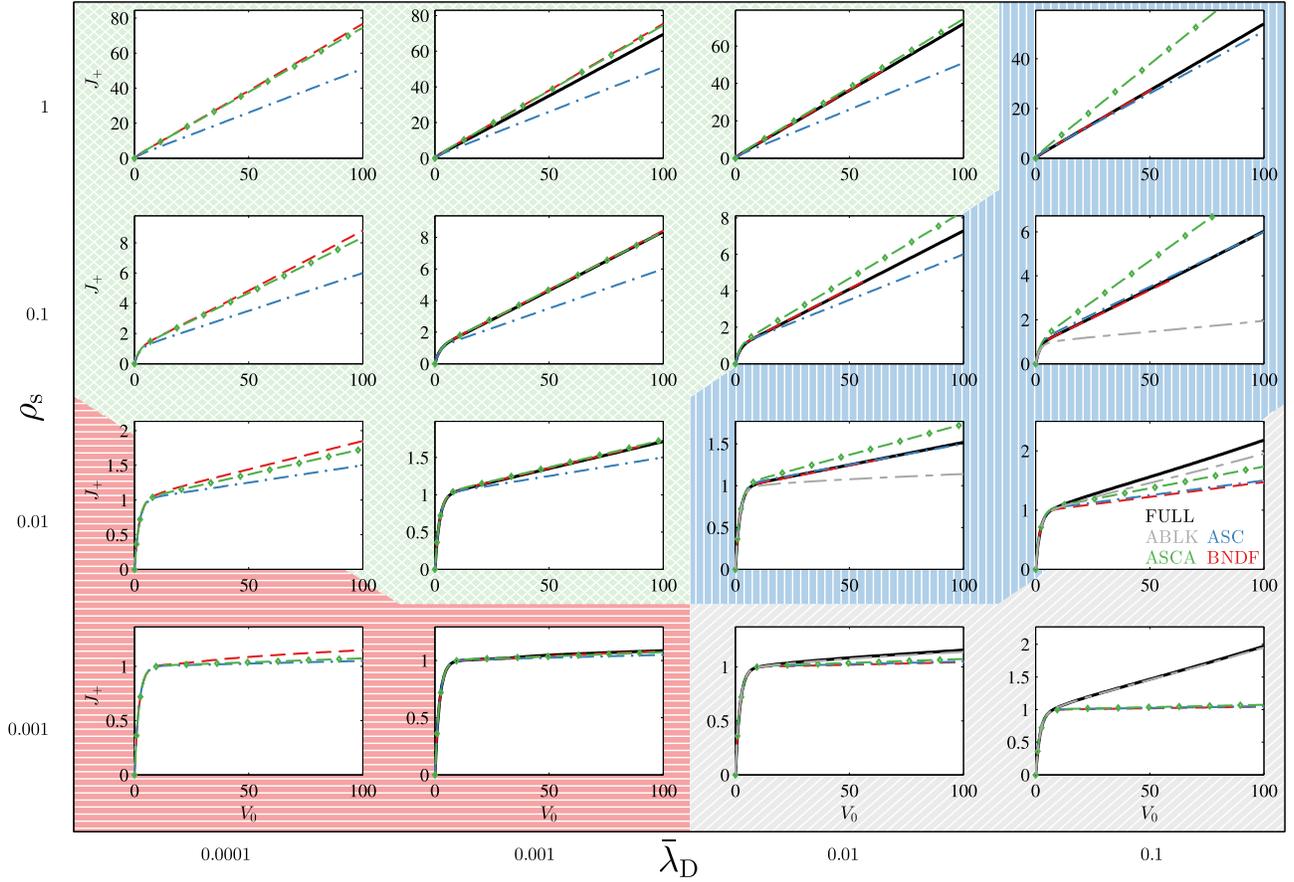}
    \caption{\figlab{IV_plots_mech_indication_alpha_05} (Color online) $I$-$V$ characteristics for $\alpha =0.05$, $\blamD=\{0.0001,0.001,0.01,0.1\}$, and $\rhos=\{0.001,0.01,0.1,1\}$. The full (black) line show the characteristics obtained from the FULL model. The dashed (red) curves are obtained from the BNDF model. The (blue) dash-dot curves are from the ASC model, and the (green) dash-diamond curves are from the ASCA model. The (gray) long-dash-short-dash curves are obtained from the ABLK model. The background patterns indicate the dominant overlimiting conduction mechanism. The (green) cross-hatched pattern indicate that surface advection and surface conduction are the dominant mechanisms. The (blue) vertically hatched pattern indicate that surface conduction without surface advection is the dominant mechanism. The (red) horizontally hatched pattern indicate that bulk advection is the dominant mechanism. The (gray) skew-hatched pattern indicate that bulk conduction through the ESC is the dominant mechanism. Intermediate cases are indicated with mixed background patterns.}
\end{figure*}

In \figsref{IV_plots_mech_indication_alpha_05}{IV_plots_mech_indication_alpha_2} the numerically calculated $I$-$V$ characteristics are plotted for a long slender channel ($\alpha=0.05$) and a short broad channel ($\alpha=0.2$), respectively. In the Supplemental Material\footnote{See Supplemental Material at [URL] for additional $I$-$V$ characteristics for $\alpha=0.01$ and $\alpha=0.1$.} additional results for $\alpha=0.01$ and $\alpha=0.1$ are given. The results for the FULL model with resolved diffuse double layers (defined in \secref{Gov_eq}) are shown in a full (black) line. The results for the BNDF model (defined in \secref{BL_model}) are shown in a dashed (red) line. The long-dash-short-dash (gray) line is obtained from the ABLK model (note that \eqref{ESC_cond_asympt} gives the asymptotic version of this curve). The dash-dot (blue) line is the analytical curve from the ASC model, and the dash-diamond (green) line is the analytical curve from the ASCA model. To help structure the results the $I$-$V$ characteristics have been given a background pattern (colored), which indicate the dominant conduction mechanisms. A light cross-hatched (green) background indicates that the dominant mechanisms are surface conduction and surface advection. A dark horizontally-hatched (red) background indicates that bulk advection is the dominant mechanism. Dark with vertical hatches (blue) indicates that surface conduction without surface advection is the dominant mechanism and light with skewed hatches (gray) indicates that the dominant mechanism is bulk conduction through the extended space-charge region. A split background indicates that the overlimiting current is the result of two different mechanisms. In the case of a split cross-hatched/vertically-hatched background, the split indicates that surface conduction is important, and that surface advection plays a role, but that this role is somewhat reduced due to backflow along the channel axis.

\begin{figure*}[!t]
    \includegraphics[width=17cm]{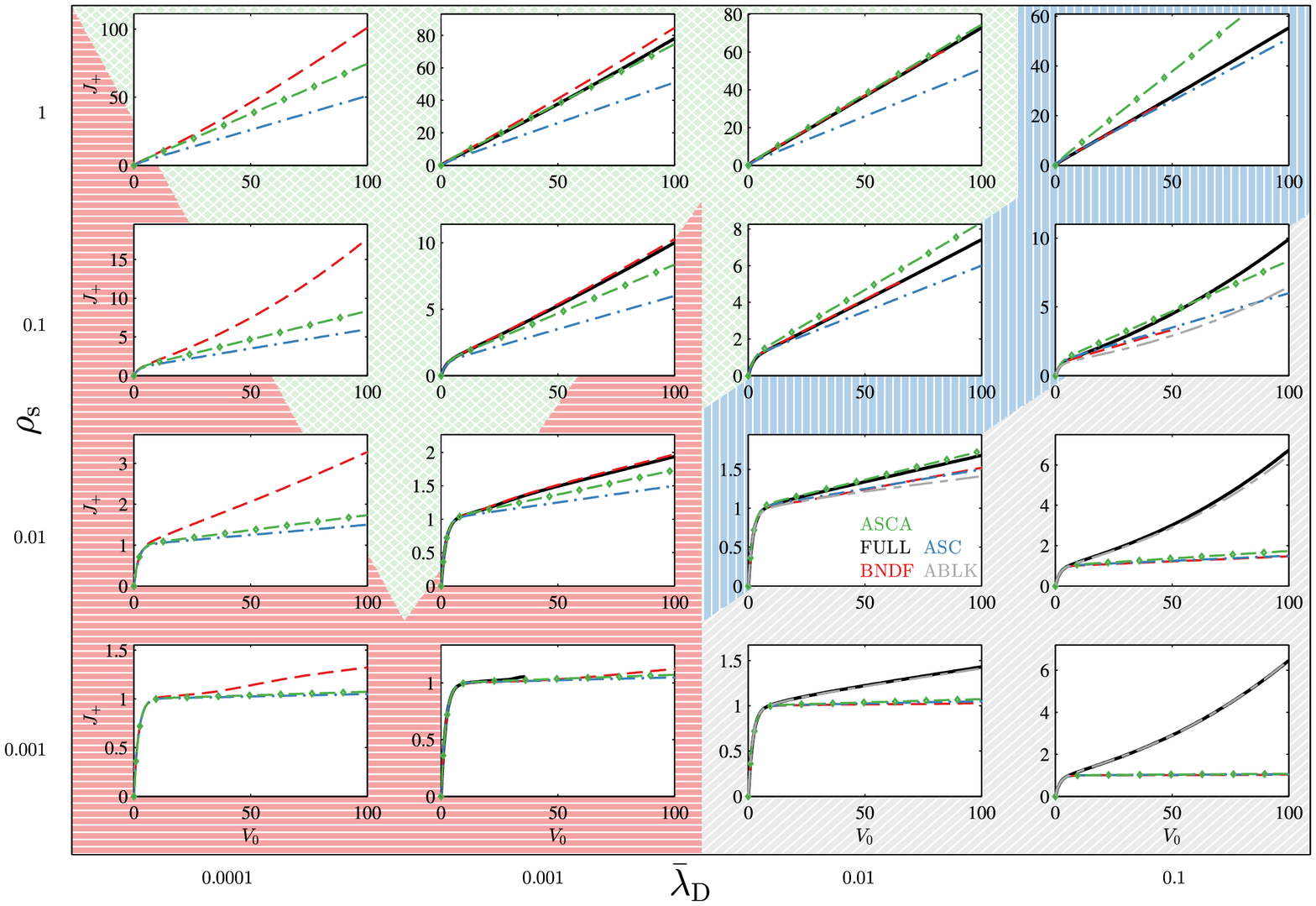}
    \caption{\figlab{IV_plots_mech_indication_alpha_2} (Color online) Same $I$-$V$ characteristics as \figref{IV_plots_mech_indication_alpha_05}, except that here $\alpha =0.2$ instead of $0.05$.}
\end{figure*}

We first consider the case $\alpha=0.05$ shown in \figref{IV_plots_mech_indication_alpha_05}. Here, the aspect ratio $\alpha$ is so low that the effects of bulk advection are nearly negligible. As a consequence the numerical (dashed [red] and full [black] lines) and analytical (dash-diamond [green] line) curves nearly match each other in a large portion of the parameter space (light cross-hatched [green] region). Although there is a small region in which bulk advection does play a role (dark horizontally-hatched [red] region), the overlimiting current due to bulk advection is small for all of the investigated $\blamD$ and $\rhos$ values. In the right part (high $\blamD$) of \figref{IV_plots_mech_indication_alpha_05} the effects of bulk and surface advection are negligible. For high $\rhos$ values surface conduction dominates (dark vertically-hatched [blue] region) and for low $\rhos$ bulk conduction through the ESC dominates (light skew-hatched [gray] region).

The case of $\alpha=0.2$, shown in \figref{IV_plots_mech_indication_alpha_2}, follows the same basic pattern as the $\alpha=0.05$ case. As expected from \figref{IV_plots_rationalize}, the regions where bulk advection (dark [red] horizontal hatches) or bulk conduction (light [gray] skewed hatches) dominates grow as $\alpha$ is increased. Inside the regions an increase in magnitude of both effects is also seen. The picture that emerges, is that in the long channel limit $\alpha \lesssim 0.05$ the effects of bulk advection are negligible, and for small $\blamD$ the overlimiting current is entirely due to surface conduction and surface advection. For bulk advection to cause a significant overlimiting current the channel has to be relatively short, $\alpha \gtrsim 0.1$, and the normalized Debye length has to be small, $\blamD \lesssim 0.001$.

\subsection{Field distributions}

In \figref{fields_plot} some of the important fields are plotted for two different sets of parameter values. The fields are obtained from the BNDF model. To the left, in panel (a), (b), and (c), the fields are given for a system with $\blamD=0.0001$, $\rhos =0.01$, $\alpha=0.2$, $V_0=60$, and to the right, in panel (d), (e), and (f), the fields are given for a system with $\blamD=0.001$, $\rhos =0.1$, $\alpha=0.05$, $V_0=60$. The colors indicate the relative magnitude (black low value, white high value) of the fields within each panel. Comparing panel (c) and (f) we see that the depletion region is bigger in panel (f) than panel (c), which is as expected since the current in panel (f) is larger than in panel (c) (cf. \figsref{IV_plots_mech_indication_alpha_05}{IV_plots_mech_indication_alpha_2}). It is also noted that the transverse distribution of the concentration is much less uniform in the (c) panel than in the (f) panel. Due to this nonuniformity (see \secref{Scalingofbulkadv}), system (a)-(b)-(c) has a net current contribution from bulk advection, whereas bulk advection contributes negligibly to the current in the transversally uniform system (d)-(e)-(f). In panel (a), we see that the majority of the current is carried in the bulk until $x\sim 0.9$, at which point it enters the boundary layer. In panel (d), on the other hand, the current enters the boundary layer already at $x\sim 0.3$, because the amount of bulk advection is insufficient to carry a bulk current into the depletion region.

\begin{figure*}[!t]
    \includegraphics[width=\textwidth]{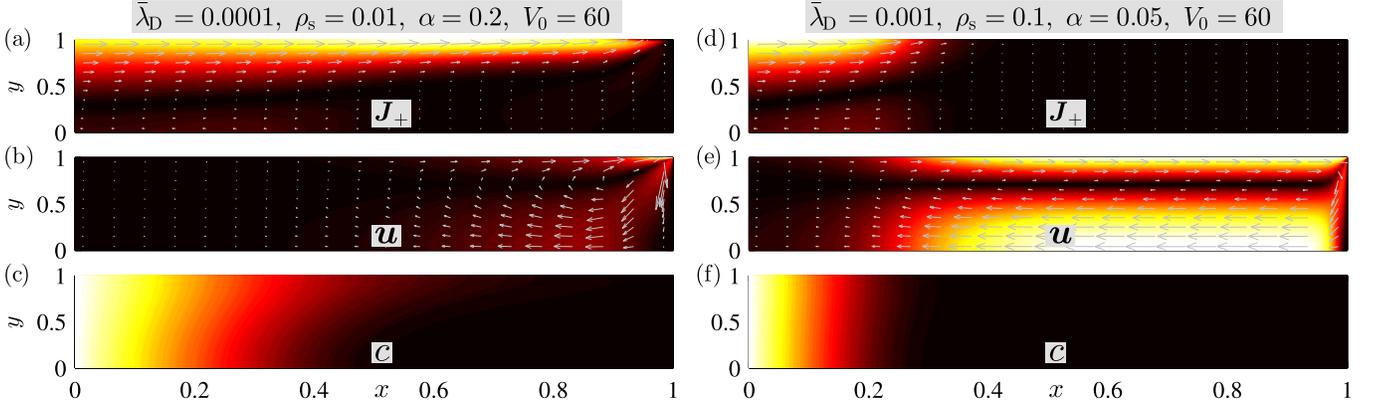}
    \caption{\figlab{fields_plot} (Color online)
    For a system with $\blamD=0.0001$, $\rhos=0.01$, $\alpha =0.2$, and $V_0=60$ is plotted (a) cation current $\vJp$, (b) velocity $\vu$, and (c) salt concentration $c$. For a system with $\blamD=0.001$, $\rhos=0.1$, $\alpha =0.05$, and $V_0=60$ is plotted (d) cation current $\vJp$, (e) velocity $\vu$, and (f) salt concentration $c$. The fields are obtained from the BNDF model, and the colors indicate the relative magnitude (black low and white high) of the fields within each panel, while arrows represent vector fields.}
\end{figure*}

\subsection{Coupling between bulk advection and the surface current}

As seen in \figsref{IV_plots_mech_indication_alpha_05}{IV_plots_mech_indication_alpha_2}, the limits of surface advection and surface conduction, of surface conduction, and of bulk conduction through the ESC, are well described by our analytical models. The analytical models do not describe the transitions between the limiting behaviours, but the essentials of the involved mechanisms are well understood. It is thus mainly the bulk advection which requires a more thorough investigation. As pointed out in Refs.~\cite{Zholkovskij2003,Zholkovskij2004, Griffiths2000,Yaroshchuk2011a}, the effects of bulk advection can to some extent be understood in terms of a Taylor--Aris-like model of hydrodynamic dispersion. However, in those papers surface conduction and surface advection is neglected on account of their small contribution to the total current in the investigated limits. It turns out that in the context of concentration polarization the surface currents do in fact play a crucial role for the bulk advection, even when the surface currents themselves only give a minute contribution to the total current. Our boundary layer model is ideally suited to demonstrate just that point, since it allows us to artificially turn off the surface currents while keeping the electro-diffusio-osmotic flow. In \figref{IV_SC_onoff}(a) $I$-$V$ characteristics obtained from the BNDF (dashed [red] line) and BNDS (dotted [purple] line) models are plotted for $\alpha=0.2$, $\blamD =0.0001$, and $\rhos =0.001$. For comparison the $I$-$V$ characteristic from the ASCA model, which includes surface conduction and surface advection but excludes bulk advection, is also plotted. In \figref{IV_SC_onoff}(b) the same curves are plotted with $\rhos =0.1$ instead of $0.001$. Comparing the BNDF model (dashed [red]) with the ASCA model (dash-diamond [green]), it is seen that bulk advection plays a significant role in these regimes. In light of this it is indeed remarkable that the BNDS model, which includes bulk advection but excludes surface currents, (dotted [purple] line) exhibits no overlimiting current at all. We conclude that the surface current is, in some way, a prerequisite for significant bulk advection.

Our investigations suggest that the reason for this highly nonlinear coupling between bulk advection and the surface current is that the surface current sets the length of the depletion region before bulk advection sets in. The large gradients in electrochemical potentials, and thereby the large electro-diffusio-osmotic velocities exist in the depletion region, so a wide depletion region implies a wide region with significant advection. In the limit of zero surface current, the depletion region only extends over a tiny region next to the membrane. In this region there is a huge electro-diffusio-osmotic flow towards the membrane, but the effects of that flow are not felt very far away, because it is compensated by the back-pressure driven flow over a quite small distance. When there is a surface current the depletion region will eventually, as the driving potential is increased, extend so far away from the membrane that back-pressure does not immediately compensate the electro-diffusio-osmotic flow. In that situation, bulk advection may begin to play a role. The need for a sufficiently large depletion region is seen by the plateau in the BNDF $I$-$V$ characteristic in \figref{IV_SC_onoff}(a). What happens is that as a function of voltage, the current increases to the limiting current, remains there for a while, and then, once the depletion region is sufficiently developed, increases further due to bulk advection. To quantify these notions we derive a simple estimate of the extent of the depletion region.

Before bulk advection sets in, the overlimiting current is entirely due to the surface current, and in this regime the behavior is well-described by the ASCA model \eqsref{sim_surf_cond_adv_phib}{sim_surf_cond_adv}. There is some ambiguity in defining exactly which parts of the system constitute the depletion region. By definition, the depletion region comprises the parts of the system, which are depleted of charge carriers. However, since there are always some charge carriers present, we have to decide on a concentration which counts as sufficiently depleted. There are a number of legitimate choices for this concentration, but for the purposes of this analysis, we define the depletion region as the part of the system where the surface conductivity exceeds the bulk conductivity. Consequently, at the beginning of the depletion region, we have from \eqref{sim_surf_cond_adv_phib1}
\begin{align}
 \ee^{\phib} = \rhos( 1+2Pe^0_+). \eqlab{Depletion_condition}
\end{align}
From \eqref{sim_surf_cond_adv}, we find the current in the overlimiting case as
\begin{align}
J_+  \approx  1 + \rhos\left (\frac{1}{2}+Pe^0_+\right )V_0,  \eqlab{Overlim_J}
\end{align}
and from \eqref{sim_surf_cond_adv_phib} the relation between position $x$ and bulk potential $\phib$ is
 \beq{JpxApprox}
 J_+ x  \approx 1-\ee^{\phib}  -\rhos(1+2Pe^0_+)\phib.
 \eeq
Inserting \eqsref{Depletion_condition}{Overlim_J} into \eqref{JpxApprox} we find the position $x_0$ where the depletion region begins
\begin{align}
x_0&= \frac{1  -\rhos(1+2Pe^0_+)\Big\{1+\ln\big[\rhos(1+2Pe^0_+)\big]\Big\}}{1 + \rhos\left (\frac{1}{2}+Pe^0_+\right )V_0}. \eqlab{Critical_V0}
\end{align}
For a small overlimiting current, the denominator is close to unity, and this implies that before bulk advection becomes important, the width $1-x_0$ of the depletion region is approximately given by
\begin{align}
1-x_0 \approx  \rhos(1+2Pe^0_+)\Big \{\frac{V_0}{2}+1+\ln\big[\rhos(1+2Pe^0_+)\big] \Big \}. \eqlab{depl_extent}
\end{align}
We can use this expression for the width of the depletion region to test our hypothesis, that the extent of the depletion region determines the onset of bulk advection. If the hypothesis is true, we should find that the overlimiting current $J_+^\mr{overlim}=J_+ - (1-\ee^{-\frac{1}{2}V_0})$ only depends on $\rhos$ and $V_0$ through the expression for $1-x_0$,
\begin{align}
J_+^\mr{overlim}(\rhos, V_0) \rightarrow J_+^\mr{overlim}( 1-x_0[\rhos,V_0]).
\end{align}
In \figref{IV_overlim_vs_V0_x0}(a) and (b) the overlimiting current $J_+^\mr{overlim}$ obtained from the BNDF model is plotted for $\rhos = \{0.0001,0.0002,0.0003,0.0004,0.0005 \}$, $\blamD=0.0001$, and $\alpha=0.05$ versus $V_0$ and $1-x_0$, respectively. The characteristic features in the curves are seen to coincide when the curves are plotted versus $1-x_0$. In contrast, no unifying behavior is seen when the curves are plotted versus $V_0$. The numerical results thus corroborate our hypothesis that the initiation of significant bulk advection is determined by the extent of the depletion region.

\begin{figure}[!t]
    \includegraphics[width=\columnwidth]{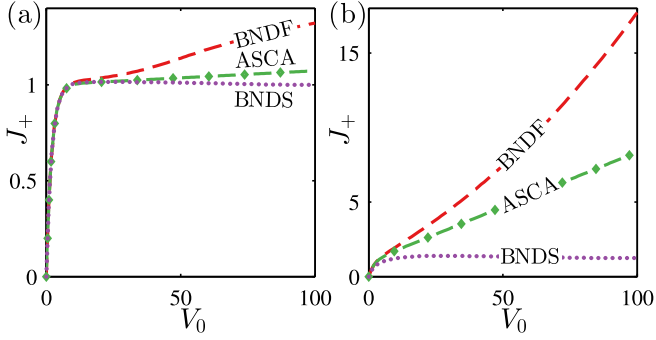}
    \caption{\figlab{IV_SC_onoff} (Color online) (a) $I$-$V$ characteristics highlighting the role of the surface current for bulk advection. $\alpha=0.2$, $\blamD=0.0001$, and $\rhos=0.001$. The dashed (red) curve is obtained from the BNDF model and the dash-diamond (green) curve is from the ASCA model. The dotted (purple) curve is obtained from the BNDS model, in which the surface current has been artificially removed while the electro-diffusio-osmotic slip velocity is kept. (b) Same as (a), but with $\rhos=0.1$.}
\end{figure}

\subsection{Issues with the numerical models}

Before concluding, we are obligated to comment on the shortcomings of the numerical models, i.e. the FULL model and the BNDF model. In the $\rhos=0.001$, $\blamD=0.001$ panel of \figref{IV_plots_mech_indication_alpha_2} the FULL model (full [black] line) is seen to break down right around $V_0\sim 40$. The reason for this breakdown is that electro-diffusio-osmosis is relatively weak and that the ESC is prone to electroosmotic instability at this $\blamD$  value. The employed steady-state model is not well-suited for modelling instabilities and therefore the model breaks down at this relatively low voltage. Because the magnitude of the ESC charge density scales as $(\alpha \blamD)^{2/3}$ we do not expect this to be an issue for the $\blamD=0.0001$ or $\alpha=0.05$ cases \cite{Nielsen2014}. Another issue seen in \figsref{IV_plots_mech_indication_alpha_05}{IV_plots_mech_indication_alpha_2}, is that in the upper right quadrant ($\rhos \geq 0.1$ and $\blamD \geq 0.01$) the BNDF model (dashed [red] curve) breaks down somewhere between $V_0\sim 40$ and $V_0 \sim 70$. The reason for this breakdown is that the Gouy length is not small in this region, as is required by the boundary layer model. The BNDF model breaks down, even though the systems in question are close to the simple transverse equilibrium configuration. The reason for this is that when the Gouy length is large, the boundary layer model underestimates the transverse transport in the system, and this eventually leads to a breakdown, when the transverse bulk transport cannot keep up with the longitudinal surface transport.

\begin{figure}[!t]
    \includegraphics[width=\columnwidth]{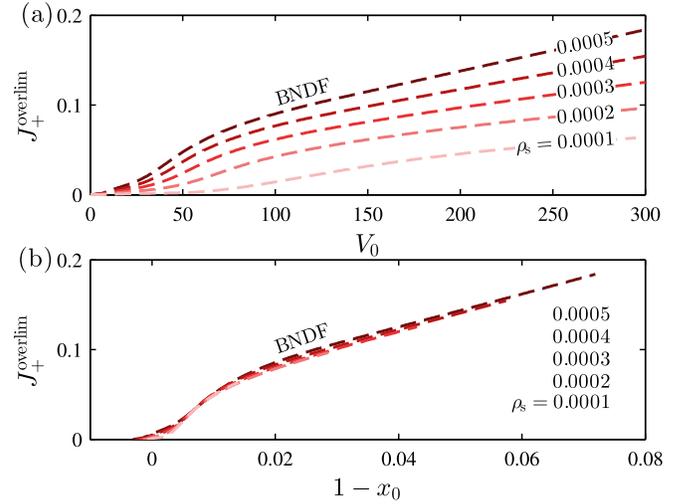}
    \caption{\figlab{IV_overlim_vs_V0_x0} (Color online) (a) The overlimiting current $J_+^\mr{overlim}$ obtained from the BNDF model for $\blamD=0.0001$, $\alpha=0.05$, and $\rhos = \{0.0001,0.0002,0.0003,0.0004,0.0005 \}$ plotted versus $V_0$. (b) Same as (a), but plotted versus $1-x_0$.   }
\end{figure}

\section{Conclusion}
\seclab{conclusion}

In this paper, we have made a thorough combined numerical and analytical study of the transport mechanisms in a microchannel undergoing concentration polarization. We have rationalized the behavior of the system and identified four mechanisms of overlimiting current: surface conduction, surface advection, bulk advection, and bulk conduction through the extended space-charge region (ESC). In the limits where surface conduction, surface advection, or bulk conduction through the ESC dominates we have derived accurate analytical models for the ion transport and verified them numerically. In the limit of long, narrow channels these models are in excellent agreement with the numerical results. We have found that bulk advection is mainly important for short, broad channels, and using numerical simulations we have quantified this notion and outlined the parameter regions with significant bulk advection. A noteworthy discovery is that the development of bulk advection is strongly dependent on the surface current, even in the cases where the surface current contributes much less to the total current than bulk advection. The numerical simulations have been carried out using both a full numerical model with resolved diffuse double layers, and an accurate boundary layer model suitable in the limit of small Gouy lengths.


%

\end{document}